\documentclass[twocolumn,preprintnumbers,amssymb,amsmath,aps,floatfix,prl,superscriptaddress]{revtex4}

\usepackage{epsfig}
\usepackage{bm}
\usepackage{amssymb}
\usepackage{amsmath}
\usepackage{color}
\usepackage{subfigure}
\usepackage{hyperref}

\begin{document}

\title{Enhancement of baryon-to-meson ratios around jets as a signature of medium response}

\author{Ao Luo}
\affiliation{Institute of Particle Physics and Key Laboratory of Quark and Lepton Physics (MOE), Central China Normal University, Wuhan, 430079, China}

\author{Ya-Xian Mao}
\affiliation{Institute of Particle Physics and Key Laboratory of Quark and Lepton Physics (MOE), Central China Normal University, Wuhan, 430079, China}

\author{Guang-You Qin}
\email{guangyou.qin@mail.ccnu.edu.cn}
\affiliation{Institute of Particle Physics and Key Laboratory of Quark and Lepton Physics (MOE), Central China Normal University, Wuhan, 430079, China}

\author{En-Ke Wang}
\email{wangek@scnu.edu.cn}
\affiliation{Guangdong Provincial Key Laboratory of Nuclear Science, Institute of Quantum Matter, South China Normal University, Guangzhou 510006, China}
\affiliation{Institute of Particle Physics and Key Laboratory of Quark and Lepton Physics (MOE), Central China Normal University, Wuhan, 430079, China}

\author{Han-Zhong Zhang}
\email{zhanghz@mail.ccnu.edu.cn}
\affiliation{Institute of Particle Physics and Key Laboratory of Quark and Lepton Physics (MOE), Central China Normal University, Wuhan, 430079, China}
\affiliation{Guangdong Provincial Key Laboratory of Nuclear Science, Institute of Quantum Matter, South China Normal University, Guangzhou 510006, China}

\begin{abstract}

We present a unique signal of jet-induced medium excitations: the enhancement of baryon-to-meson ratios around the quenched jets.
To illustrate this, we study jet-particle correlations and the distributions of jet-induced identified particles with respect to the jet direction in Pb+Pb collisions at the LHC via a multi-phase transport model.
We find a strong enhancement of baryon-to-meson ratios for associated particles at intermediate transverse momentum around the triggered jets in Pb+Pb collisions relative to p+p collisions, due to the coalescence of jet-excited medium partons.
Since the lost energy from jets can diffuse to large angles, such baryon-to-meson-ratio enhancement is more pronounced for larger relative distance from the jet axis.
We argue that the experimental confirmation of the enhancement of jet-induced baryon-to-meson ratios around the jets will provide an unambiguous evidence for the medium response to jet quenching in heavy-ion collisions.

\end{abstract}
\maketitle

{\it Introduction.} Jet quenching provides one of the most remarkable evidences for the discovery of quark-gluon plasma (QGP) in high-energy heavy-ion collisions \cite{Wang:1991xy, Qin:2015srf, Majumder:2010qh, Blaizot:2015lma, Cao:2020wlm}.
The interaction between jets and medium can degrade the energy of jet partons and change the energy and momentum distributions among jet constituents.
Extensive jet quenching studies \cite{JET:2013cls, JETSCAPE:2021ehl, Xing:2019xae, Huss:2020dwe, Zhao:2021vmu, Qin:2010mn, Majumder:2011uk, Dai:2012am, Blaizot:2013hx, Chien:2016led, Caucal:2018dla, Mehtar-Tani:2021fud} have revealed rich information about the properties of QGP produced at the Relativistic Heavy-Ion Collider (RHIC) and the Large Hadron Collider (LHC).
Collaborative efforts have found that jet transport coefficient in the hot QGP formed at RHIC and the LHC is about two orders of magnitude larger than that in a cold nucleus \cite{JET:2013cls, JETSCAPE:2021ehl}.

Jet-medium interaction can also lead to medium excitations, such as the Mach cone induced by supersonic partons propagating through the QGP \cite{CasalderreySolana:2004qm, Stoecker:2004qu, Chaudhuri:2005vc, Ruppert:2005uz, Gubser:2007ga, Chesler:2007an, Qin:2009uh, Neufeld:2009ep, Li:2010ts}, which can in turn affect the dynamical evolution of bulk matter and various jet-related observables \cite{Ma:2010dv, Tachibana:2014lja, Floerchinger:2014yqa, Andrade:2014swa, Gao:2016ldo, Brewer:2017fqy, Milhano:2017nzm, Yan:2017rku, Chang:2019sae, Tachibana:2020mtb, Casalderrey-Solana:2020rsj}.
Detailed studies of medium excitations can probe much information about the equation of state and other bulk properties, such as shear viscosity, of the QCD medium \cite{Neufeld:2008dx, Bouras:2014rea}.
Unfortunately, the direct detection of jet-induced flow and Mach cone in heavy-ion collisions is extremely difficult since the collective flow of the rapidly expanding medium can significantly distort the Mach cone structure \cite{Ma:2010dv, Betz:2010qh, Tachibana:2015qxa}.

Tremendous efforts have been devoted to investigate the medium response effect and search for its signal.
Reference \cite{Tachibana:2017syd} finds that jet-induced medium flow and excitations can diffuse to large angles, and may dominate jet shape at large radius.
Similar effects have also been found in Refs. \cite{KunnawalkamElayavalli:2017hxo, Luo:2018pto, Park:2018acg}.
Recently, CMS Collaboration \cite{CMS:2016cvr, CMS:2016qnj, CMS:2018zze, CMS:2021nhn} has measured the distribution of charged particles around the quenched jets and found a significant enhancement of low-$p_T$ particles, which extends large relative angles of $\Delta \eta \approx 1$ and $\Delta \phi \approx 1$. This is consistent with the expectation that the energy lost at high $p_T$ due to jet-medium interaction reappears in the form of low-$p_T$ particles far away from the jet axis \cite{Tachibana:2017syd, Luo:2021hoo}.
References \cite{Chen:2017zte, Chen:2020tbl} find that jet-induced medium response can also lead to the enhancement of soft particles in $\gamma$-hadron correlations and $\gamma$-jet fragmentation functions.
Reference \cite{Pablos:2019ngg} studies the medium response effect on jet suppression from small and large radius.
Very recently, Ref. \cite{Chen:2021adl} proposes to use the two-dimensional jet tomography technique to locate the initial jet positions to help detect the diffusion wake in $Z/\gamma$-jet events.
It is still an on-going effort to search for the signal of jet-induced medium response in relativistic heavy-ion collisions.
Identification and isolation of medium excitations are of paramount importance for understanding the details of jet-medium interaction and the bulk properties and evolution dynamics of QGP.

This work presents a unique signature of medium response: the enhancement of jet-induced baryon-to-meson yield ratios around the jets.
We note that the enhanced baryon-to-meson ratios at intermediate $p_T$ region in heavy-ion collisions have been successfully explained by the coalescence of partons in the bulk matter, which is also the key to interpret the number of constituent quark scaling of elliptic flow, another important evidence for the QGP formation \cite{Fries:2003vb, Fries:2003kq, Molnar:2003ff, Greco:2003xt, Greco:2003mm, Greco:2003vf, Hwa:2004ng, STAR:2004jwm, STAR:2005npq, PHENIX:2006dpn, STAR:2007zea, PHENIX:2012swz}.
Similarly, the lost energy from jets is deposited to medium partons, which, via parton coalescence, can change the chemical composition of particles such as the relative yields of baryons to mesons around the quenched jets as compared to the vacuum jets.
To illustrate this, we use jet-particle correlation method to compute the ($\Delta \eta, \Delta \phi$) distribution of jet-induced baryon and meson yields with respect to the jet direction in Pb+Pb and p+p collisions at $\sqrt{s_{NN}} = 5.02$~TeV at the LHC via the AMPT model.
The mixed-event and side-band methods are applied to compute the background-subtracted jet-induced identified particle yields.
We find a strong enhancement of jet-induced baryon-to-meson yield ratios for associated particles at intermediate $p_T$ around the triggered jets in Pb+Pb collisions relative to p+p collisions, as a result of the coalescence of medium partons which are excited by the hard jets.
This enhancement is more prominent in more central collisions due to larger jet quenching effect.
Since the lost energy from jets can diffuse to large angles, the jet-triggered baryon-to-meson ratios in Pb+Pb collisions are more enhanced for larger relative distance $\Delta r = \sqrt{(\Delta\eta)^2 + (\Delta\phi)^2}$ from the jet axis  \footnote{The proceeding \cite{Chen:2021rrp} studies the effect of jet-induced medium excitation on $\Lambda/K_S^0$ ratio in jets, whereas we focus on the baryon-to-meson ratios around jets.
Since the lost energy from jets can flow to large angles, the signal of medium response is much stronger outside jets than inside jets.
Additionally,
we use jet-particle correlation method which provides a systematic procedure to subtract the background and extract jet-induced identified particle yields up to very large distance from the jet direction.
}.
Our finding can be easily tested by experiments using the same jet-particle correlation method.
Once confirmed, it will provide an unambiguous evidence for jet-induced medium exciations in heavy-ion collisions.

{\it Jet-particle correlations.} We use the AMPT model \cite{Lin:2004en, Zhang:2005ni, Ma:2010dv} to simulate the dynamical evolution of bulk matter and jet-medium interaction in relativistic heavy-ion collisions.
This coupled model has been used for describing many bulk and jet observables, such as anisotropic collective flows \cite{Lin:2001zk, Chen:2004dv, Xu:2011fe, Ma:2010dv}, dijet and $\gamma$-jet asymmetries \cite{Ma:2013pha, Ma:2013bia}, jet fragmentation function \cite{Ma:2013gga} and jet shape \cite{Ma:2013uqa}.
The AMPT model contains four main stages: initial condition, parton cascade, hadronization and hadron cascade.
In the string-melting version of the AMPT model, hadrons produced from the HIJING model \cite{Wang:1991hta, Gyulassy:1994ew} are first converted to their valence quarks and antiquarks.
To generate jet events with high statistics, a jet trigger technique in HIJING is applied in the AMPT model.
Then, the dynamical evolution of medium partons together with jet partons is simulated via ZPC parton cascade model \cite{Zhang:1997ej}, in which only elastic parton scatterings are included.
In this work, we set the partonic cross section as $1.5$~mb to study Pb+Pb collisions at $\sqrt{s_{NN}} = 5.02$~TeV at the LHC.
When the partonic system freezes out, a quark coalescence model is used to combine partons into hadrons \cite{Lin:2001zk}.
Then the produced hadrons evolve via the ART model \cite{Li:1995pra}, which includes elastic and inelastic baryon-baryon, baryon-meson and meson-meson scatterings.

To compute jet-induced particle yields around the quenched jets, we use the correlations between the triggered jets and the associated identified particles. Such jet-particle correlation method has been used by CMS Collaboration \cite{CMS:2016cvr, CMS:2016qnj, CMS:2018zze, CMS:2021nhn} and also in our previous work \cite{Luo:2021hoo} to study the distribution of charged particles around jets.
Here we focus on jet-induced baryon and meson productions.
First, we construct the two-dimensional identified particle number density $\frac{d^2N}{d\Delta\phi d\Delta\eta}$ per trigger jet, where ${\Delta \eta}$ and ${\Delta \phi}$ are relative pseudo-rapidity and azimuthal angle with respect to the jet direction.
In this work, we take the reconstructed jets with $p^{\rm jet}_{T} > 120$ GeV, $R = 0.4$ and $| \eta_{\rm jet} | < 1.6$ and identified particles with $|\eta| < 2.4$.
To correct the limited acceptance effect on jet-particle correlations, we use a mixed-event method.
Then the acceptance-corrected two-dimensional identified particle density per trigger jet can be obtained \cite{CMS:2016cvr, CMS:2016qnj, CMS:2018zze, CMS:2021nhn}:
\begin{align}
	\frac{1}{N_{\rm jet}} \frac{d^2N}{d{\Delta\eta} d{\Delta\phi}}
	= \frac{ME(0,0)}{ME({\Delta \eta},{\Delta\phi})} S({\Delta \eta},{\Delta\phi}).
\label{eq:dN_per_jet}
\end{align}
Here, the signal pair distribution $S({\Delta \eta},{\Delta \phi})$ and the mixed-event pair distribution $ME({\Delta \eta},{\Delta \phi})$ represent the per-trigger-jet yields of associated particles from the same and mixed events,
\begin{align}
	S({\Delta \eta},{\Delta\phi}) &= \frac{1}{N_{\rm jet}} \frac{d^2N^{\rm same}}{d{\Delta\eta} d{\Delta\phi}},
\nonumber\\
	ME({\Delta \eta},{\Delta\phi}) & =
	\frac{1}{N_{\rm jet}} \frac{d^2N^{\rm mixed}}{d{\Delta\eta} d{\Delta\phi}}.
\end{align}
The ratio $\frac{ME(0,0)}{ME({\Delta \eta},{\Delta\phi})}$ is the normalized correction factor.

\begin{figure}
	\includegraphics[width=0.99\linewidth]{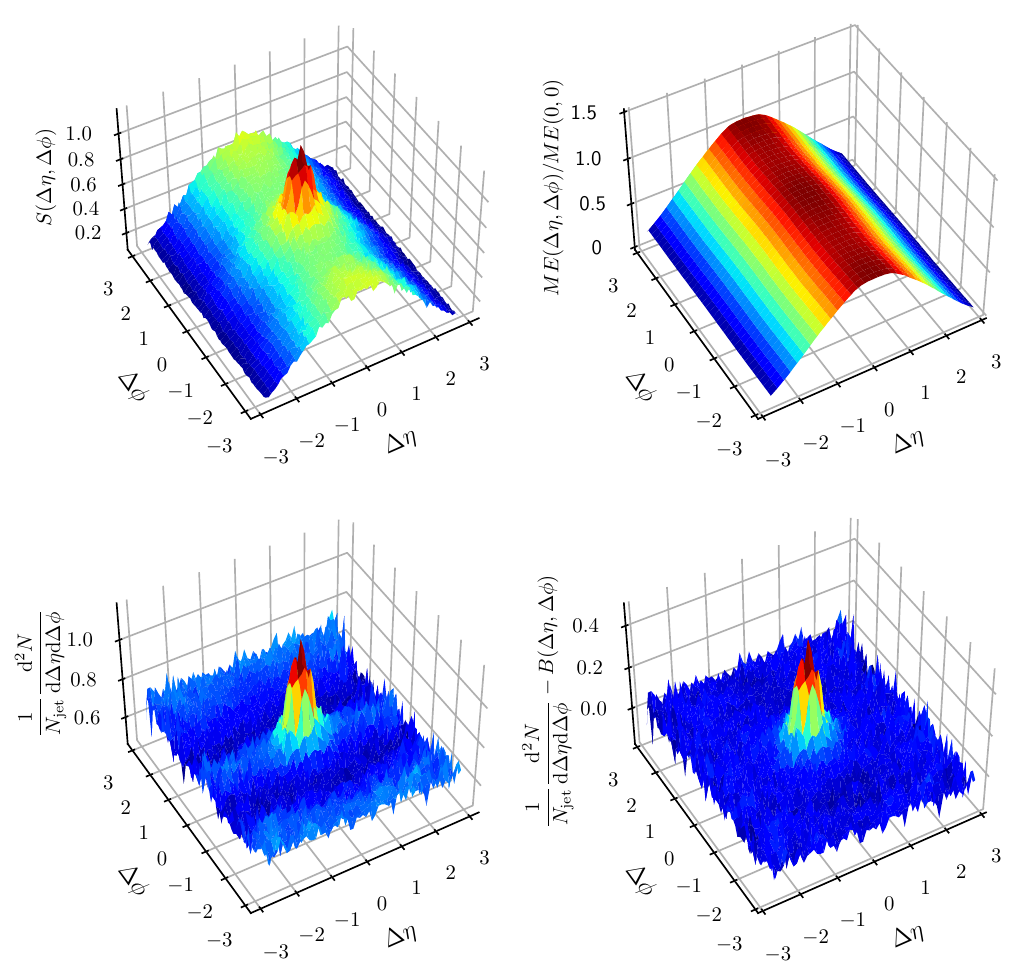}
	\caption{Jet-proton correlations for associated proton $p_T = 2-3$~GeV in central $0-10\%$ Pb+Pb collisions. The upper left is the signal pair distribution, the upper right is the normalized mixed-event pair distribution, the lower left is the acceptance-corrected distribution, and the lower right is the final background-subtracted jet-triggered proton yield. }
	\label{fig:mixed_event}
\end{figure}

\begin{figure*}
	\includegraphics[width=0.99\linewidth]{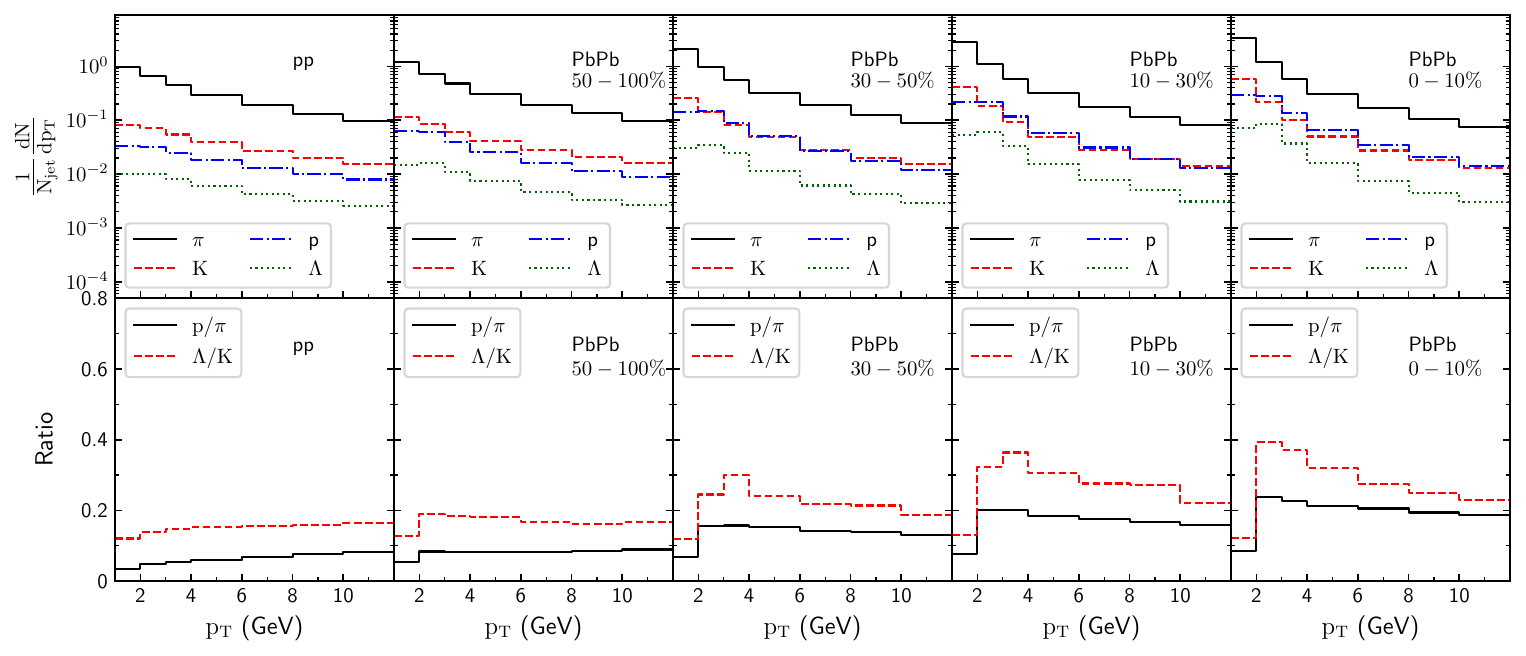}
	\caption{Jet-induced identified particle yields in the region ${\Delta r<1}$ around the jets as a function of associated particle $p_T$ for p+p and $50-100\%, 30-50\%, 10-30\%, 0-10\%$ Pb+Pb collisions at $\sqrt{s_{NN}} = 5.02$~TeV (upper panels). The lower panels show the corresponding $p/\pi$ and $\Lambda/K$ ratios as a function of associated particle $p_T$.
Jets are taken with ${p^{\rm jet}_T} > 120$ GeV, $R = 0.4$ and $| \eta_{\rm jet} | < 1.6$.
}
	\label{fig:dNdpT}
\end{figure*}

To illustrate the above procedure, we show in Fig. \ref{fig:mixed_event} the two-dimensional distributions of the associated protons with $2 < p_T <3$ GeV around the triggered jets in $0-10\%$ Pb+Pb collisions.
We first use the anti-$k_t$ algorithm in the FASTJET package to reconstruct the triggered jets \cite{Cacciari:2008gp, Cacciari:2011ma}.
Then we use the associated protons around the triggered jets to construct the signal pair distribution $S(\Delta \eta, \Delta \phi)$, as shown in the upper left panel of Fig. \ref{fig:mixed_event}.
One can see a jet-like peak on the near side and a much smeared peak on the away side.
Next, we construct the mixed-event distribution $ME(\Delta \eta, \Delta \phi)$ by pairing the triggered jet and particles from other events.
The upper right panel shows the normalized mixed-event pair distribution $ME(\Delta \eta, \Delta \phi)/ME(0,0)$.
The lower left panel shows the acceptance-corrected distribution for associated protons per trigger jet.
Finally, we apply the side-band method to subtract the background contribution from uncorrelated pairs and long-range correlations.
Following CMS \cite{CMS:2016qnj}, we estimate the background contribution using $\Delta \phi$ distribution averaged over $1.5 < |\Delta \eta| < 2.5$.
The final corrected jet-proton correlations are shown in the lower right panel of Fig. \ref{fig:mixed_event}.

The above two-dimensional number density distribution $\frac{d^2N}{d\Delta\phi d\Delta\eta}$ can be studied for different $p_T$ of identified particles.
In this work, we divide the identified particles into several $p_T$ bins bounded by 1, 2, 3, 4, 6, 8, 10, 12~GeV.
In this sense, we have constructed the three-dimensional distribution $\frac{d^3N}{dp_T d\Delta \eta d\Delta \phi}$ for identified particles around the quenched jets.
From this distribution, one can study the $p_T$ distributions of identified particles around the reconstructed jets by integrating out the rapidity and angular parts ($\Delta \eta$ and $\Delta \phi$):
\begin{equation}
	\frac{dN}{dp_T} = \int d\Delta \phi \int d\Delta \eta
	\frac{d^3 N}{dp_T d\Delta \phi d\Delta \eta } \bigg| _{\Delta r < 1}.
\end{equation}
Here we look at identified particles in the region with $\Delta r < 1$ around the jet axis.
One may also study the radial $\Delta r$ distribution of identified particles around jets, which can be obtained as follows:
\begin{align}
	\frac{dN}{d\Delta r} &= \int d\Delta \phi \int d\Delta \eta \int dp_T
	\frac{d^3 N}{dp_T d\Delta \phi d\Delta \eta }
\nonumber\\
& \times \delta(\Delta r - \sqrt{(\Delta\phi)^2 + (\Delta\eta)^2}).
\end{align}
In practice, the above radial distribution is constructed in annular rings of width $0.1$ for $\Delta r$ around the jet direction.

\begin{figure}
\includegraphics[width=0.81\linewidth]{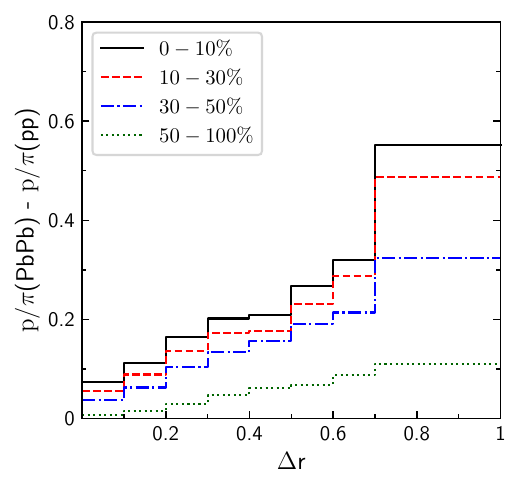}
\includegraphics[width=0.81\linewidth]{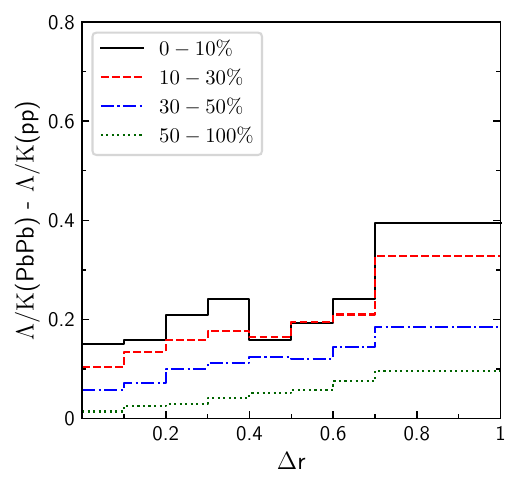}
	\caption{The enhancement of $p/\pi$ and $\Lambda/K$ ratios around the jets for associated particle $p_T =2 - 6$~GeV as a function of $\Delta r$ in 0-10\%, 10-30\%, 30-50\% and 50-100\% Pb+Pb collisions at $\sqrt{s_{NN}} = 5.02$~TeV relative to p+p collisions.
Jets are taken with ${p^{\rm jet}_T} > 120$ GeV, $R = 0.4$ and $| \eta_{\rm jet} | < 1.6$.
	}
	\label{fig:AA-pp_Deltar}
\end{figure}

{\it Jet-induced baryon and meson yields.} In Fig.~\ref{fig:dNdpT}, we show jet-induced $\pi$, $K$, $p$ and $\Lambda$ yields $\frac{1}{N_{\rm jet}}\frac{dN}{dp_T}$ in the region ${\Delta r<1}$ around the triggered jets as a function of particle $p_T$ for p+p and Pb+Pb collisions at $\sqrt{s_{NN}} = 5.02$~TeV, as calculated from the AMPT model (upper panels).
The results for Pb+Pb collisions are shown for four centralities ($50-100\%, 30-50\%, 10-30\%, 0-10\%$).
To increase the statistics, we average over $\pi^+$, $\pi^-$ and $\pi^0$ and for brevity use $\pi$ to denote $\frac{1}{3}(\pi^+ + \pi^- + \pi^0)$.
Similarly, we use $p$ for the average $\frac{1}{2}(p + \bar{p})$, $K$ for $\frac{1}{3}(K^+ + K^- + K_S^0)$, and $\Lambda$ for the average $\frac{1}{2}(\Lambda + \bar{\Lambda})$.
One can clearly see an enhancement of low $p_T$ particles in Pb+Pb collisions compared to p+p collisions.
This means that a significant fraction of the lost energy from jets is carried by soft particles at large angles with respect to the jet direction \cite{Tachibana:2017syd, Luo:2021hoo, CMS:2016qnj, CMS:2018zze, ATLAS:2019pid}, consistent with experimental results \cite{CMS:2016qnj, CMS:2018zze, ATLAS:2019pid}.
The lower panels show the corresponding $p/\pi$ and $\Lambda/K$ ratios in the region ${\Delta r<1}$ around the jets as a function of $p_T$.
It is interesting to note that the ratios of $\Lambda/K$ around jets in p+p collisions from the AMPT model are comparable to the ALICE measurement of $\Lambda/K_S^0$ ratio in jets, despite different collision and jet energies \cite{ALICE:2021cvd}.
A very important result is the strong enhancement of jet-induced $p/\pi$ and $\Lambda/K$ ratios for associated particles at intermediate $p_T$ ($2-6$~GeV) around the jets in Pb+Pb collisions compared to p+p collisions.
Such baryon-to-meson-ratio enhancement at intermediate $p_T$ region is a general signature of parton coalescence \cite{Fries:2003vb, Fries:2003kq, Molnar:2003ff, Greco:2003xt, Greco:2003mm, Greco:2003vf, Hwa:2004ng}.
Since the lost energy from jets is deposited to medium partons, the production of baryons relative to mesons at intermediate $p_T$ region gets more enhanced via parton coalescence.
One can clearly see that the enhancement is more pronounced in more central collisions in which jet quenching effect is larger.
This means that the enhancement of jet-induced baryon-to-meson ratios for associated particles at intermediate $p_T$ around the jets is mainly caused by the medium response to jet quenching.
We note that since AMPT only includes parton coalescence, the associated particle yields at high $p_T$ region is not reliable. For $p_T > 6-8$~GeV, one needs to consider more dominant contribution from fragmentation, which is left to future effort.

Now we focus on the intermediate $p_T$ regime and look at the radial $\Delta r$ dependence of the baryon-to-meson enhancement around the quenched jets. In Fig.~\ref{fig:AA-pp_Deltar}, we show the enhancement of $p/\pi$ and $\Lambda/K$ ratios around the triggered jets for associated particle $p_T=2-6$~GeV as a function of the relative distance $\Delta r$ with respect to the jet direction for Pb+Pb collisions at $\sqrt{s_{NN}} = 5.02$~TeV relative to p+p collisions.
Note that in order to improve the poor statistics at large $\Delta r$, we have combined four sub bins together to get the last bin $\Delta r = 0.6-1$.
One can clearly see that the enhancement of the baryon-to-meson ratios around jets is more pronounced in more central collisions due to larger jet quenching effect.
Another very interesting observation is that the enhancement of jet-induced baryon-to-meson ratios is stronger for larger distance $\Delta r$.
Such radial dependence can be understood because the lost energy from the quenched jets can diffuse to large angles.
Due to parton coalescence, the production of baryons relative to mesons at intermediate $p_T$ is more enhanced at larger distance with respect to the jet direction.
Note that AMPT does not include radiative process which may affect the energy deposition by jets into the medium and jet-induced medium response \cite{Qin:2009uh, Neufeld:2009ep}.
Therefore, the above illustration is only qualitative.
However, since the baryon-to-meson enhancement is a general feature of parton coalescence, regardless the details of the coalescence models, our qualitative prediction of such unique signature of medium response should be robust.

{\it Summary.} We have presented a unique signature of the medium response to jet quenching: the enhancement of baryon-to-meson ratios around the quenched jets.
Using the jet-particle correlation method, we have studied how jet-induced medium excitations change the final-state chemical compositions of the particles produced around the quenched jets in heavy-ion collisions.
In particular, we have computed the ($\Delta \eta, \Delta \phi$) distribution of baryons and mesons with respect to the triggered jet direction in Pb+Pb collisions and p+p collisions at $\sqrt{s_{NN}} = 5.02$~TeV at the LHC via the AMPT model.
In order to extract jet-induced identified particle yields around the jets, we have applied the mixed-event and side-band methods to subtract the contributions from uncorrelated background and long-range correlations.
Our result shows that jet-induced baryon-to-meson yield ratios for associated particles at intermediate $p_T$ around the quenched jets are significantly enhanced in Pb+Pb collisions compared to p+p collisions, as a result of the coalescence of jet-excited medium partons.
Such enhancement is more pronounced in more central Pb+Pb collisions due to stronger jet-medium interaction effect.
We further show that the enhancement of jet-triggered baryon-to-meson ratios for associated particles at intermediate $p_T$ is stronger for larger relative distance $\Delta r$ from the jet direction since the lost energy from jets can flow to large angles.
Our findings can be easily tested by the experiments.
Once confirmed, it will provide an unambiguous evidence for the medium response to jet quenching in heavy-ion collisions.
A very interesting future direction is to apply our method to proton-nucleus collisions and other small systems, which may provide a sensitive tool to detect the signal of jet quenching in small systems.

{\it Acknowledgments.} We thank S. Cao, B. M\"uller and X.-N. Wang for useful discussions. This work is supported in part by National Natural Science Foundation of China under Grants Nos. 12225503, 11775095, 11890710, 11890711, 11935007 and 11975005, and the Guangdong Major Project of Basic and Applied Basic Research (No. 2020B0301030008).

\bibliographystyle{h-physrev5} 
\bibliography{refs_GYQ}   

\begin{thebibliography}{10}

\bibitem{Wang:1991xy}
X.-N. Wang and M.~Gyulassy,
\newblock Phys.Rev.Lett. {\bf 68}, 1480 (1992).

\bibitem{Qin:2015srf}
G.-Y. Qin and X.-N. Wang,
\newblock Int. J. Mod. Phys. E {\bf 24}, 1530014 (2015), arXiv:1511.00790.

\bibitem{Majumder:2010qh}
A.~Majumder and M.~Van~Leeuwen,
\newblock Prog. Part. Nucl. Phys. {\bf 66}, 41 (2011), arXiv:1002.2206.

\bibitem{Blaizot:2015lma}
J.-P. Blaizot and Y.~Mehtar-Tani,
\newblock Int. J. Mod. Phys. {\bf E24}, 1530012 (2015), arXiv:1503.05958.

\bibitem{Cao:2020wlm}
S.~Cao and X.-N. Wang,
\newblock Rept. Prog. Phys. {\bf 84}, 024301 (2021), arXiv:2002.04028.

\bibitem{JET:2013cls}
JET, K.~M. Burke {\em et~al.},
\newblock Phys. Rev. C {\bf 90}, 014909 (2014), arXiv:1312.5003.

\bibitem{JETSCAPE:2021ehl}
JETSCAPE, S.~Cao {\em et~al.},
\newblock Phys. Rev. C {\bf 104}, 024905 (2021), arXiv:2102.11337.

\bibitem{Xing:2019xae}
W.-J. Xing, S.~Cao, G.-Y. Qin, and H.~Xing,
\newblock Phys. Lett. B {\bf 805}, 135424 (2020), arXiv:1906.00413.

\bibitem{Huss:2020dwe}
A.~Huss {\em et~al.},
\newblock Phys. Rev. Lett. {\bf 126}, 192301 (2021), arXiv:2007.13754.

\bibitem{Zhao:2021vmu}
W.~Zhao, W.~Ke, W.~Chen, T.~Luo, and X.-N. Wang,
\newblock (2021), arXiv:2103.14657.

\bibitem{Qin:2010mn}
G.-Y. Qin and B.~Muller,
\newblock Phys. Rev. Lett. {\bf 106}, 162302 (2011), arXiv:1012.5280,
\newblock [Erratum: Phys. Rev. Lett.108,189904(2012)].

\bibitem{Majumder:2011uk}
A.~Majumder and C.~Shen,
\newblock Phys. Rev. Lett. {\bf 109}, 202301 (2012), arXiv:1103.0809.

\bibitem{Dai:2012am}
W.~Dai, I.~Vitev, and B.-W. Zhang,
\newblock Phys. Rev. Lett. {\bf 110}, 142001 (2013), arXiv:1207.5177.

\bibitem{Blaizot:2013hx}
J.-P. Blaizot, E.~Iancu, and Y.~Mehtar-Tani,
\newblock Phys.Rev.Lett. {\bf 111}, 052001 (2013), arXiv:1301.6102.

\bibitem{Chien:2016led}
Y.-T. Chien and I.~Vitev,
\newblock Phys. Rev. Lett. {\bf 119}, 112301 (2017), arXiv:1608.07283.

\bibitem{Caucal:2018dla}
P.~Caucal, E.~Iancu, A.~H. Mueller, and G.~Soyez,
\newblock Phys. Rev. Lett. {\bf 120}, 232001 (2018), arXiv:1801.09703.

\bibitem{Mehtar-Tani:2021fud}
Y.~Mehtar-Tani, D.~Pablos, and K.~Tywoniuk,
\newblock Phys. Rev. Lett. {\bf 127}, 252301 (2021), arXiv:2101.01742.

\bibitem{CasalderreySolana:2004qm}
J.~Casalderrey-Solana, E.~V. Shuryak, and D.~Teaney,
\newblock J. Phys. Conf. Ser. {\bf 27}, 22 (2005), arXiv:hep-ph/0411315.

\bibitem{Stoecker:2004qu}
H.~Stoecker,
\newblock Nucl. Phys. A {\bf 750}, 121 (2005), arXiv:nucl-th/0406018.

\bibitem{Chaudhuri:2005vc}
A.~K. Chaudhuri and U.~Heinz,
\newblock Phys. Rev. Lett. {\bf 97}, 062301 (2006), arXiv:nucl-th/0503028.

\bibitem{Ruppert:2005uz}
J.~Ruppert and B.~Muller,
\newblock Phys. Lett. B {\bf 618}, 123 (2005), arXiv:hep-ph/0503158.

\bibitem{Gubser:2007ga}
S.~S. Gubser, S.~S. Pufu, and A.~Yarom,
\newblock Phys. Rev. Lett. {\bf 100}, 012301 (2008), arXiv:0706.4307.

\bibitem{Chesler:2007an}
P.~M. Chesler and L.~G. Yaffe,
\newblock Phys. Rev. Lett. {\bf 99}, 152001 (2007), arXiv:0706.0368.

\bibitem{Qin:2009uh}
G.~Y. Qin, A.~Majumder, H.~Song, and U.~Heinz,
\newblock Phys. Rev. Lett. {\bf 103}, 152303 (2009), arXiv:0903.2255.

\bibitem{Neufeld:2009ep}
R.~B. Neufeld and B.~Muller,
\newblock Phys. Rev. Lett. {\bf 103}, 042301 (2009), arXiv:0902.2950.

\bibitem{Li:2010ts}
H.~Li, F.~Liu, G.-l. Ma, X.-N. Wang, and Y.~Zhu,
\newblock Phys. Rev. Lett. {\bf 106}, 012301 (2011), arXiv:1006.2893.

\bibitem{Ma:2010dv}
G.-L. Ma and X.-N. Wang,
\newblock Phys. Rev. Lett. {\bf 106}, 162301 (2011), arXiv:1011.5249.

\bibitem{Tachibana:2014lja}
Y.~Tachibana and T.~Hirano,
\newblock Phys. Rev. C {\bf 90}, 021902 (2014), arXiv:1402.6469.

\bibitem{Floerchinger:2014yqa}
S.~Floerchinger and K.~C. Zapp,
\newblock Eur. Phys. J. C {\bf 74}, 3189 (2014), arXiv:1407.1782.

\bibitem{Andrade:2014swa}
R.~P.~G. Andrade, J.~Noronha, and G.~S. Denicol,
\newblock Phys. Rev. {\bf C90}, 024914 (2014), arXiv:1403.1789.

\bibitem{Gao:2016ldo}
Z.~Gao, A.~Luo, G.-L. Ma, G.-Y. Qin, and H.-Z. Zhang,
\newblock Phys. Rev. C {\bf 97}, 044903 (2018), arXiv:1612.02548.

\bibitem{Brewer:2017fqy}
J.~Brewer, K.~Rajagopal, A.~Sadofyev, and W.~Van Der~Schee,
\newblock JHEP {\bf 02}, 015 (2018), arXiv:1710.03237.

\bibitem{Milhano:2017nzm}
G.~Milhano, U.~A. Wiedemann, and K.~C. Zapp,
\newblock Phys. Lett. B {\bf 779}, 409 (2018), arXiv:1707.04142.

\bibitem{Yan:2017rku}
L.~Yan, S.~Jeon, and C.~Gale,
\newblock Phys. Rev. C {\bf 97}, 034914 (2018), arXiv:1707.09519.

\bibitem{Chang:2019sae}
N.-B. Chang, Y.~Tachibana, and G.-Y. Qin,
\newblock Phys. Lett. B {\bf 801}, 135181 (2020), arXiv:1906.09562.

\bibitem{Tachibana:2020mtb}
Y.~Tachibana, C.~Shen, and A.~Majumder,
\newblock (2020), arXiv:2001.08321.

\bibitem{Casalderrey-Solana:2020rsj}
J.~Casalderrey-Solana, J.~G. Milhano, D.~Pablos, K.~Rajagopal, and X.~Yao,
\newblock JHEP {\bf 05}, 230 (2021), arXiv:2010.01140.

\bibitem{Neufeld:2008dx}
R.~Neufeld,
\newblock Phys.Rev. {\bf C79}, 054909 (2009), arXiv:0807.2996.

\bibitem{Bouras:2014rea}
I.~Bouras, B.~Betz, Z.~Xu, and C.~Greiner,
\newblock Phys.Rev. {\bf C90}, 024904 (2014), arXiv:1401.3019.

\bibitem{Betz:2010qh}
B.~Betz, J.~Noronha, G.~Torrieri, M.~Gyulassy, and D.~H. Rischke,
\newblock Phys. Rev. Lett. {\bf 105}, 222301 (2010), arXiv:1005.5461.

\bibitem{Tachibana:2015qxa}
Y.~Tachibana and T.~Hirano,
\newblock Phys. Rev. C {\bf 93}, 054907 (2016), arXiv:1510.06966.

\bibitem{Tachibana:2017syd}
Y.~Tachibana, N.-B. Chang, and G.-Y. Qin,
\newblock Phys. Rev. C {\bf 95}, 044909 (2017), arXiv:1701.07951.

\bibitem{KunnawalkamElayavalli:2017hxo}
R.~Kunnawalkam~Elayavalli and K.~C. Zapp,
\newblock JHEP {\bf 07}, 141 (2017), arXiv:1707.01539.

\bibitem{Luo:2018pto}
T.~Luo, S.~Cao, Y.~He, and X.-N. Wang,
\newblock Phys. Lett. B {\bf 782}, 707 (2018), arXiv:1803.06785.

\bibitem{Park:2018acg}
C.~Park, S.~Jeon, and C.~Gale,
\newblock Nucl. Phys. A {\bf 982}, 643 (2019), arXiv:1807.06550.

\bibitem{CMS:2016cvr}
CMS, V.~Khachatryan {\em et~al.},
\newblock JHEP {\bf 11}, 055 (2016), arXiv:1609.02466.

\bibitem{CMS:2016qnj}
CMS, V.~Khachatryan {\em et~al.},
\newblock JHEP {\bf 02}, 156 (2016), arXiv:1601.00079.

\bibitem{CMS:2018zze}
CMS, A.~M. Sirunyan {\em et~al.},
\newblock JHEP {\bf 05}, 006 (2018), arXiv:1803.00042.

\bibitem{CMS:2021nhn}
CMS, A.~M. Sirunyan {\em et~al.},
\newblock JHEP {\bf 05}, 116 (2021), arXiv:2101.04720.

\bibitem{Luo:2021hoo}
A.~Luo, Y.-X. Mao, G.-Y. Qin, E.-K. Wang, and H.-Z. Zhang,
\newblock (2021), arXiv:2107.11751.

\bibitem{Chen:2017zte}
W.~Chen, S.~Cao, T.~Luo, L.-G. Pang, and X.-N. Wang,
\newblock Phys. Lett. B {\bf 777}, 86 (2018), arXiv:1704.03648.

\bibitem{Chen:2020tbl}
W.~Chen, S.~Cao, T.~Luo, L.-G. Pang, and X.-N. Wang,
\newblock Phys. Lett. B {\bf 810}, 135783 (2020), arXiv:2005.09678.

\bibitem{Pablos:2019ngg}
D.~Pablos,
\newblock Phys. Rev. Lett. {\bf 124}, 052301 (2020), arXiv:1907.12301.

\bibitem{Chen:2021adl}
W.~Chen {\em et~al.},
\newblock Phys. Rev. Lett. {\bf 127}, 082301 (2021), arXiv:2101.05422.

\bibitem{Fries:2003vb}
R.~J. Fries, B.~Muller, C.~Nonaka, and S.~A. Bass,
\newblock Phys. Rev. Lett. {\bf 90}, 202303 (2003), arXiv:nucl-th/0301087.

\bibitem{Fries:2003kq}
R.~J. Fries, B.~Muller, C.~Nonaka, and S.~A. Bass,
\newblock Phys. Rev. C {\bf 68}, 044902 (2003), arXiv:nucl-th/0306027.

\bibitem{Molnar:2003ff}
D.~Molnar and S.~A. Voloshin,
\newblock Phys. Rev. Lett. {\bf 91}, 092301 (2003), arXiv:nucl-th/0302014.

\bibitem{Greco:2003xt}
V.~Greco, C.~M. Ko, and P.~Levai,
\newblock Phys. Rev. Lett. {\bf 90}, 202302 (2003), arXiv:nucl-th/0301093.

\bibitem{Greco:2003mm}
V.~Greco, C.~M. Ko, and P.~Levai,
\newblock Phys. Rev. C {\bf 68}, 034904 (2003), arXiv:nucl-th/0305024.

\bibitem{Greco:2003vf}
V.~Greco, C.~M. Ko, and R.~Rapp,
\newblock Phys. Lett. B {\bf 595}, 202 (2004), arXiv:nucl-th/0312100.

\bibitem{Hwa:2004ng}
R.~C. Hwa and C.~B. Yang,
\newblock Phys. Rev. C {\bf 70}, 024905 (2004), arXiv:nucl-th/0401001.

\bibitem{STAR:2004jwm}
STAR, J.~Adams {\em et~al.},
\newblock Phys. Rev. C {\bf 72}, 014904 (2005), arXiv:nucl-ex/0409033.

\bibitem{STAR:2005npq}
STAR, J.~Adams {\em et~al.},
\newblock Phys. Rev. Lett. {\bf 95}, 122301 (2005), arXiv:nucl-ex/0504022.

\bibitem{PHENIX:2006dpn}
PHENIX, A.~Adare {\em et~al.},
\newblock Phys. Rev. Lett. {\bf 98}, 162301 (2007), arXiv:nucl-ex/0608033.

\bibitem{STAR:2007zea}
STAR, B.~I. Abelev {\em et~al.},
\newblock Phys. Lett. B {\bf 655}, 104 (2007), arXiv:nucl-ex/0703040.

\bibitem{PHENIX:2012swz}
PHENIX, A.~Adare {\em et~al.},
\newblock Phys. Rev. C {\bf 85}, 064914 (2012), arXiv:1203.2644.

\bibitem{Chen:2021rrp}
W.~Chen, S.~Cao, T.~Luo, L.-G. Pang, and X.-N. Wang,
\newblock Nucl. Phys. A {\bf 1005}, 121934 (2021).

\bibitem{Lin:2004en}
Z.-W. Lin, C.~M. Ko, B.-A. Li, B.~Zhang, and S.~Pal,
\newblock Phys. Rev. C {\bf 72}, 064901 (2005), arXiv:nucl-th/0411110.

\bibitem{Zhang:2005ni}
B.~Zhang, L.-W. Chen, and C.-M. Ko,
\newblock Phys. Rev. C {\bf 72}, 024906 (2005), arXiv:nucl-th/0502056.

\bibitem{Lin:2001zk}
Z.-w. Lin and C.~M. Ko,
\newblock Phys. Rev. C {\bf 65}, 034904 (2002), arXiv:nucl-th/0108039.

\bibitem{Chen:2004dv}
L.-W. Chen, C.~M. Ko, and Z.-W. Lin,
\newblock Phys. Rev. C {\bf 69}, 031901 (2004), arXiv:nucl-th/0312124.

\bibitem{Xu:2011fe}
J.~Xu and C.~M. Ko,
\newblock Phys. Rev. C {\bf 84}, 014903 (2011), arXiv:1103.5187.

\bibitem{Ma:2013pha}
G.-L. Ma,
\newblock Phys. Rev. C {\bf 87}, 064901 (2013), arXiv:1304.2841.

\bibitem{Ma:2013bia}
G.-L. Ma,
\newblock Phys. Lett. B {\bf 724}, 278 (2013), arXiv:1302.5873.

\bibitem{Ma:2013gga}
G.-L. Ma,
\newblock Phys. Rev. C {\bf 88}, 021902 (2013), arXiv:1306.1306.

\bibitem{Ma:2013uqa}
G.-L. Ma,
\newblock Phys. Rev. C {\bf 89}, 024902 (2014), arXiv:1309.5555.

\bibitem{Wang:1991hta}
X.-N. Wang and M.~Gyulassy,
\newblock Phys. Rev. D {\bf 44}, 3501 (1991).

\bibitem{Gyulassy:1994ew}
M.~Gyulassy and X.-N. Wang,
\newblock Comput. Phys. Commun. {\bf 83}, 307 (1994), arXiv:nucl-th/9502021.

\bibitem{Zhang:1997ej}
B.~Zhang,
\newblock Comput. Phys. Commun. {\bf 109}, 193 (1998), arXiv:nucl-th/9709009.

\bibitem{Li:1995pra}
B.-A. Li and C.~M. Ko,
\newblock Phys. Rev. C {\bf 52}, 2037 (1995), arXiv:nucl-th/9505016.

\bibitem{Cacciari:2008gp}
M.~Cacciari, G.~P. Salam, and G.~Soyez,
\newblock JHEP {\bf 04}, 063 (2008), arXiv:0802.1189.

\bibitem{Cacciari:2011ma}
M.~Cacciari, G.~P. Salam, and G.~Soyez,
\newblock Eur. Phys. J. C {\bf 72}, 1896 (2012), arXiv:1111.6097.

\bibitem{ATLAS:2019pid}
ATLAS, G.~Aad {\em et~al.},
\newblock Phys. Rev. C {\bf 100}, 064901 (2019), arXiv:1908.05264,
\newblock [Erratum: Phys.Rev.C 101, 059903 (2020)].

\bibitem{ALICE:2021cvd}
ALICE, S.~Acharya {\em et~al.},
\newblock (2021), arXiv:2105.04890.

\end{thebibliography}

\end{document}